\documentclass[11pt]{article}
\usepackage{iap2000,epsfig}
\def\kms{km~s$^{-1}$}
\def\lta{\mathrel{\rlap{\lower 3pt\hbox{$\mathchar"218$}}
     \raise 2.0pt\hbox{$\mathchar"13C$}}}
\def\gta{\mathrel{\rlap{\lower 3pt\hbox{$\mathchar"218$}}
     \raise 2.0pt\hbox{$\mathchar"13E$}}}
\def\etal{{et~al.}} 
\def\sn{\ifmmode S_N\else$S_N$\fi}
\newcommand\hkpc{$h^{-1}\,$kpc}
\newcommand\mo{\ifmmode m^0\else$m^0$\fi}
\def\hst{{\it HST}}

\def\aj{AJ}
\def\apj{ApJ}
\def\apj{ApJ}
\def\apjl{ApJ}

\def\mnras{MNRAS}

\def\araa{ARA\&A}
\def\pasp{PASP}

\begin{document}
\vspace*{4cm}
\title{HUDDLED MASSES YEARNING TO STREAM FREE: \\
GLOBULAR CLUSTERS IN THE HEARTS OF GALAXY CLUSTERS}

\author{JOHN P. BLAKESLEE}

\address{Department of Physics, University of Durham,
South Road, Durham, DH1 3LE, United Kingdom}

\maketitle\abstracts{%
Extremely rich populations of globular clusters, numbering ten
thousand or more, surround the central giant galaxies in rich clusters.
I discuss some recent spectroscopic and photometric observations of
these rich globular clusters populations.
In very nearby galaxy clusters, the globular cluster velocities can be used to
trace the transition from the galaxy potential to the larger cluster
potential.  Beyond $c{z}{\,\approx\,}3000$ \kms, the globulars are too faint
for spectroscopic studies with even the largest telescopes and cease to be
useful dynamical tracers.  However, the spatial distribution and size of
the globular population also contains information on the central mass 
distribution of the cluster and can be measured out to
$c{z}{\,\gta\,}20$,000 \kms.
Overall, the data suggest that these rich globular cluster populations are
comprised of normal, old globulars, the majority of which 
were assembled in the cluster centers before most of the stars in the
cD galaxy were formed.   }


\section{Introduction: Globular Clusters and their Uses}\label{sec:intro}

Globular clusters (GCs) provide key information for a wide variety of
areas in astrophysics and cosmology.  Absolute GC ages (see reviews
by Salaris \etal\ [1997] and Chaboyer \etal\ [1999]) constrain the
cosmological model by setting a lower bound on the age of the universe.
The observed GC age range (e.g., Stetson \etal\ 1996; Sarajedini \etal\
1997) constrains the time-scale of galaxy formation, and the distribution
in metallicity can provide further details on the formation processes
(e.g., Ashman \& Zepf 1992; Ajhar \etal\ 1994; Forbes \etal\ 1997).

Globular clusters are also an important part of the distance scale.  Their
horizontal-branch magnitude is the basis for the ``Population~II'' RR~Lyra
distances within our Galaxy (e.g., Carney \etal\ 1992; Carretta \etal\
2000) and provides an important self-consistency check (e.g., Walker 1992;
Ajhar \etal\ 1996; Fusi Pecci \etal\ 1996) of the Cepheid distance scale
on which most extragalactic distances are based.  
In addition, the mean luminosity
of the GCs in external galaxies has been used extensively as a standard-candle
distance indicator reaching as far as the Coma cluster
(e.g., Harris 1991; Baum \etal\ 1995, 1997; Whitmore \etal\ 1995;
Kavelaars \etal\ 2000; Ferrarese \etal\ 2000).

Besides these other, more traditional, ``uses,'' globular clusters are
now recognized as unique and important probes of the centers of 
galaxy clusters.  The present paper summarizes some recent work
on this topic.

\section{Globular Clusters as Kinematical Tracers of the Cluster Potential}
\label{sec:kin}

The existence of massive dark halos around spiral galaxies has been known
for over 20 years (Rubin \etal\ 1978) from their flat rotation curves.
The presence of a dominant dark matter component in galaxy clusters has
been known even longer from the galaxy dynamics (Zwicky 1937), and
repeatedly confirmed by X-ray, lensing, and other dynamical studies.  Yet,
massive dark halos around individual giant ellipticals have been notoriously
difficult to demonstrate and measure.  The stellar velocity dispersion can
only be traced to $\sim\,10$~kpc, roughly 20\% of the distance to which spiral
rotation curves can be followed, and then it becomes necessary to 
resort to various, usually sparse, tracers of the halo potential. 
For the same reason, the mass distribution in cluster cores is poorly
constrained at radii $r\approx10$--50~kpc.

A major step forward came with the work of Cohen \& Ryzhov (1997), who
used the 10-m Keck telescope
to measure velocities for 205 confirmed GCs around M87, the
giant elliptical at the dynamical center of the Virgo cluster.  Velocities
for an additional 16 GCs were added by Cohen (2000).  These authors found
that the 1-d velocity dispersion of the M87 GC population rises from
$\sigma\approx300$ \kms\ at $r\approx6$ kpc to $\sigma>450$ \kms\ at
$r>30$ kpc (assuming the Virgo Cepheid distance of 16 Mpc, so that
$1^\prime = 4.7$ kpc). Interestingly, the rotation also
increases outward from the center, 
reaching $v_{\rm rot}\approx 300$ \kms\ (Cohen 2000).
Cohen \& Ryzhov concluded that the total mass increases as
$M({<}r)\sim r^{1.7}$, or $\rho\sim r^{-1.3}$, which agrees well  
with the mass profile inferred from X-ray data 
(Nulsen \& B\"{o}hringer 1995) in the region of overlap.  A similar result
was found by Romanowsky \& Kochanek (2000), who determined the mass
profile using joint constraints from the GCs and halo stars and
concluded that the potential of the Virgo cluster is dominant by
$r\approx20$ kpc.  Thus, the GC velocity data trace the transition from
the dynamics of the M87 stellar halo to that of surrounding Virgo cluster.

Kissler-Patig \etal\ (1999) studied a sample of 74 GC velocities around
NGC\,1399, the central cD galaxy in the Fornax cluster ($d\approx19$ Mpc),
using data drawn from three previous spectroscopic studies.
They also found a rising velocity dispersion profile, bridging the
``velocity gap'' between the NGC\,1399 stellar halo
and the galaxy population of the Fornax cluster.  These authors
concluded that most of the extended population of GCs were stripped
from other Fornax galaxies, as advocated by
C\^ot\'e \etal\ (1998).

Recently, a team led by R.~Sharples at Durham
and S.~Zepf at Yale (Sharples \etal\ 1997; Zepf \etal\ 2000)
and another from Caltech (J.~Cohen, P.~C\^ot\'e, and
myself) have obtained velocity data for large samples of GCs around
NGC\,4472.  Although NGC\,4472 is not a central cluster galaxy, it is the
brightest Virgo member and the center of its own smaller subcluster
$\sim\,$1\,Mpc away from M87 in projection.  It has also been the subject
of a very detailed photometric study using Washington photometry
(Geisler \etal\ 1996).  The GC system contains two unusually well
demarcated metallicity subpopulations, as compared
to most other giant ellipticals (e.g., Gebhardt \& Kissler-Patig 1999;
Woodworth \& Harris 2000; Brodie \etal~2000).

\begin{figure}[h]
\centering
\psfig{figure=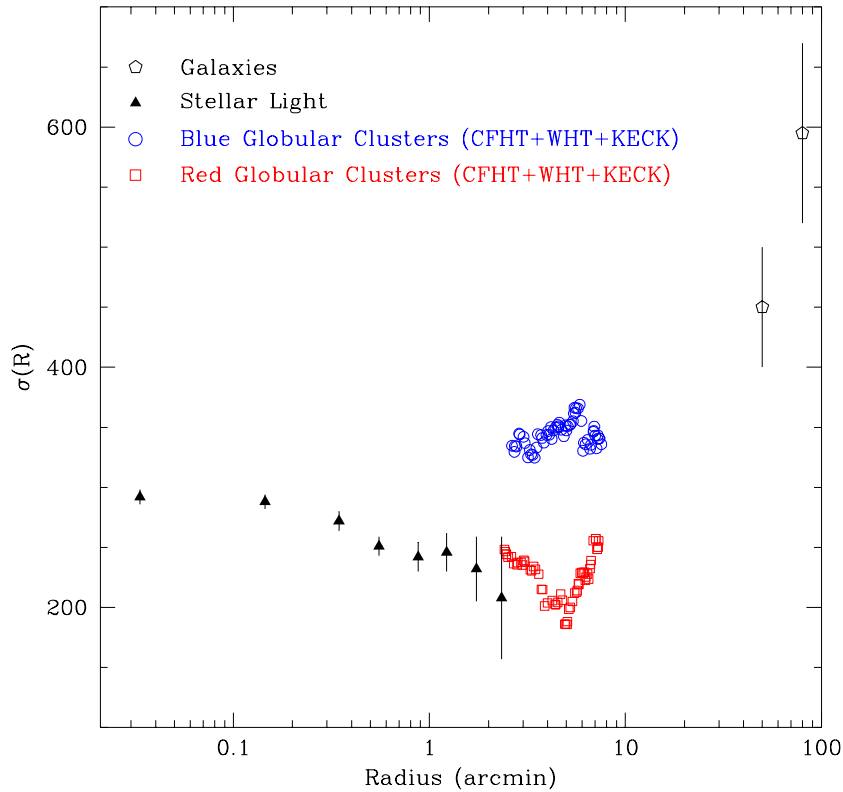,height=4.5in}
\caption{A comparison of the velocity dispersion profiles for 166~blue and
106~red globular clusters in NGC\,4472 (data from the Caltech and Durham
groups) to that of the stellar light (from Fisher \etal\ 1995) and
satellite galaxies (from NED).  The data points for the GCs are not
all independent but show a heavily over-sampled running average to produce
a smooth profile.  While the red GC population is kinematically
(as well as spatially and chemically) similar to the stellar halo,
the more spatially-extended blue GC population traces the transition
\hbox to 6.5in{from the galaxy to the cluster potential.\hfil}\label{fig:m49}}
\end{figure}

Figure~\ref{fig:m49} compares the velocity dispersion data
for 272 NGC\,4472 GCs, split into blue and red subpopulations following
Geisler \etal\ (1996), to that of the stellar halo and the satellite
galaxy population.  The red GCs share the kinematics of, and have colors
similar to, the stellar halo, bespeaking a common origin.  The blue GCs
have a significantly higher dispersion and show the transition to the
dynamics of galaxy population.  The blue population is also more extended
in its spatial distribution.  However, both subpopulations appear
equally old, within the uncertainties (Beasley \etal\ 2000), and the
same is true in M87 (Cohen \etal\ 1998).

Although it appears universally true that blue GC populations have
more spatially extended distributions, it is unclear whether or not the
striking difference in the dynamics of the two subpopulations is a common
feature of the GCs around central cluster galaxies.  There is no evidence
for this effect in the available M87 GC data (Cohen \etal\ 1998), but the
sample was not so clearly separated into blue and red components.  In
NGC\,1399, the sample size is too small in to constitute a reliable test.
Further spectroscopic observations in these two important
systems are needed.  Velocity measurements for the brightest GCs around 
the Hydra cluster cD NGC\,3311, a giant galaxy with a very
diffuse cD halo and a broad GC color distribution lacking any
apparent multiple peaks (Brodie \etal~2000), would be extremely
valuable and is now feasible with the VLT.

\section{Globular Cluster Number as a Mass Tracer}

At distances much beyond the Virgo and Fornax clusters, it becomes
increasingly difficult to measure accurate velocities for individual
GCs. For instance, at $c{z}\sim3500$ \kms, the very brightest GCs have an
apparent magnitude $V\sim22.6$, requiring very long integrations to
obtain adequate signal in the absorption line spectra.  However,
abundant GC populations can be photometrically
detected around bright galaxies to many
times this distance, and their properties contain useful information on
the host galaxies and galaxy clusters.

\subsection{Measuring GC Specific Frequencies}

The globular cluster ``specific frequency'' \sn\ of a galaxy is the number
of GCs per unit $M_V{\,=\,}{-}15$ of galaxy luminosity (Harris \& van den
Bergh 1981).  Thus, a dwarf galaxy with $M_V{\,=\,}{-}$15 and a single GC would
have $\sn{\,=\,}1$.  Spiral galaxies such as the Milky Way or M31 typically
have $\sn=0.5$--1, while most ellipticals have $\sn=2$--4.  However, cD
galaxies such as M87 can have $\sn\approx12$, which means a total GC
population of $N_{\rm GC}\approx12$,500 (see the review by Harris 1991).
However, data on other such ``high-\sn'' systems were very sparse, as the
total number of GCs is difficult to estimate from GC counts if the data do
not approach the turnover magnitude \mo\ of the globular cluster
luminosity function (GCLF).  At the distance of the Coma cluster, 
the GCLF has a $V$-band turnover $m^0_V\approx27.5$.

To address this problem, we (Blakeslee \& Tonry 1995)
developed a new method for simultaneously measuring specific frequencies 
and GCLF widths in relatively distant galaxies without the requirement
of complete point source detection to the GCLF turnover.
The method borrows from the surface brightness fluctuations (SBF)
method (Tonry \& Schneider 1988) for measuring galaxy distances. 
In the standard SBF technique, the fluctuations produced by the
Poisson statistics of stars in an early-type galaxy are
used to derive an average flux for the stellar population, yielding 
the distance (see the recent review by Blakeslee, Ajhar, \& Tonry 1999).
Here, we apply the same analysis methods to measure the fluctuations
from unresolved GCs around more distant galaxies.

\begin{figure}[h]
\centering{\psfig{figure=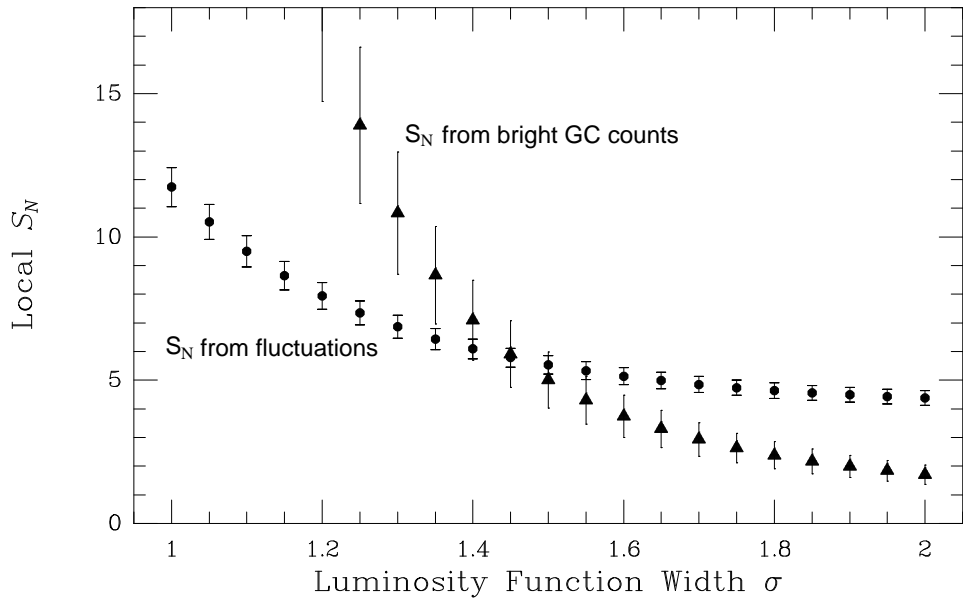, width=5.0in}}
\caption{An illustration of the joint \sn-GCLF constraints using our
count-fluctuation method.  The figure shows \sn\ values for the Coma cD
NGC\,4874 at a radius $r\sim1^{\prime}$, as derived from point-source
counts (triangles) and the power-spectrum
of the surface brightness distribution (``fluctuations'', circles)
after removal of the detected point sources.
Both measurements must be corrected for background contamination to yield the
contributions from bright and faint GCs, respectively.
Plotted errorbars include only measurement error and background
uncertainty.
The counts are much more sensitive to the assumed GCLF width $\sigma$
because they represent only the bright tail of the luminosity 
distribution.  
The two types of measurement yield consistent values of 
$S_N\approx6$ for $\sigma\approx 1.45$ mag
\hbox to 6.5in{(data from Blakeslee \& Tonry 1995).\hfil}
\label{fig:method}}
\end{figure}

The usual way of determining $S_N$ in galaxies beyond Virgo 
is to count the brightest GCs and extrapolate, yielding a result which
is strongly dependent upon the assumed GCLF width~$\sigma$.
In our technique, we also first identify and count the brightest GCs,
determining $S_N$ as a function of $\sigma$, but
we then excise these bright ones and measure the
psf-convolved variance (``fluctuations'') in the image from the rest of
the GC population.  This variance is directly proportional
to the number density of remaining GCs; thus, we again determine $S_N$ as
function of $\sigma$.  These two determinations will be consistent
over some range in $\sigma$ which should encompass the true value.  
Figure~\ref{fig:method} provides an illustration.
Even if there is relatively little variation in the GCLF $\sigma$
among cD galaxies, the combined information from the counts of the 
brightest GCs and the variance from the fainter ones produces
much tighter constraints on the total GC number.

\subsection{Ground-based Suveys and Interpretations}

\begin{figure}
\centering{\psfig{figure=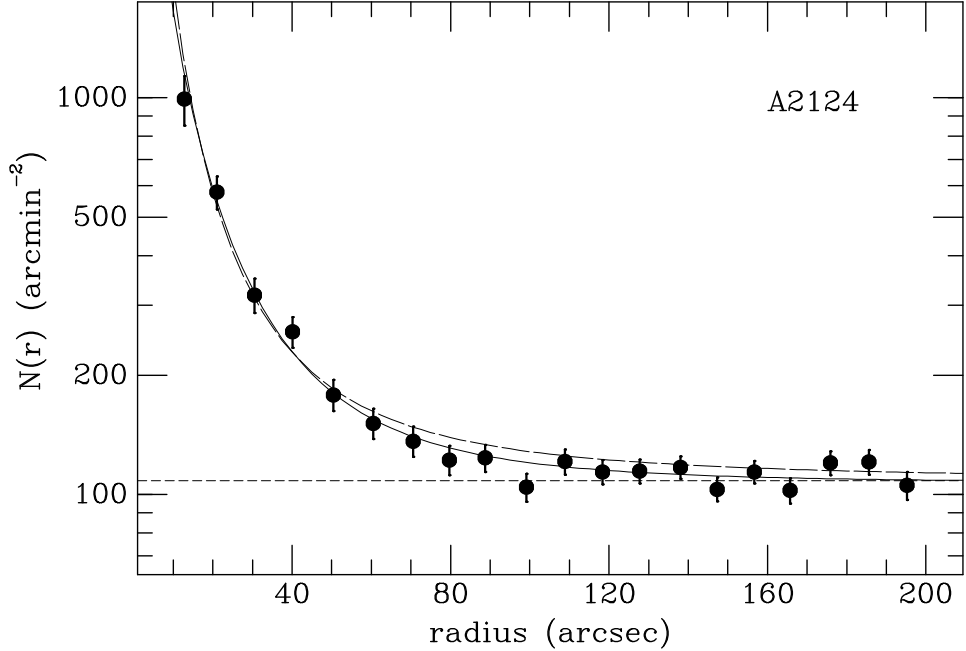, width=5.0in}}
\caption{The incompleteness-corrected surface
density of point sources in the extinction-corrected R~magnitude range
$23.6\le m_R \le26.6$ around the central cD galaxy in A2124
at $z{\,=\,}0.066$ ($1^{\prime} = 52$ \hkpc)  (Blakeslee 1999).
The solid and long-dashed curves represent de~Vaucouleurs $r^{{1/4}}$~law 
and power-law fits, respectively, while the short-dashed line
is the estimated background.
The GC population is detected to at least $r\approx150\,$kpc~($h{\,=\,}0.7$).
\label{fig:a2124rad}}
\end{figure}

We studied two different samples of GC populations in Abell cluster
central galaxies using this new method.  The first was a statistically
complete sample of 19 northern clusters with early-type brightest members
within $c{z}<10,000$ \kms\ observed with the 2.4-m telescope at MDM
Observatory on Kitt Peak (Blakeslee \etal\ 1997).  The second was a
smaller sample of more massive clusters to twice this distance with Keck
(Blakeslee 1999).  In all, \sn\ values were measured for 32 bright
galaxies in 25 clusters.  Figure~\ref{fig:a2124rad} shows data for the
most distant cluster studied, A2124 at $z{=}0.066$, which also happens to
be the nearest cluster in which strong gravitational lensing of distant
galaxy has been observed (Blakeslee \& Metzger 1999)

\begin{figure}
\centerline{\psfig{figure=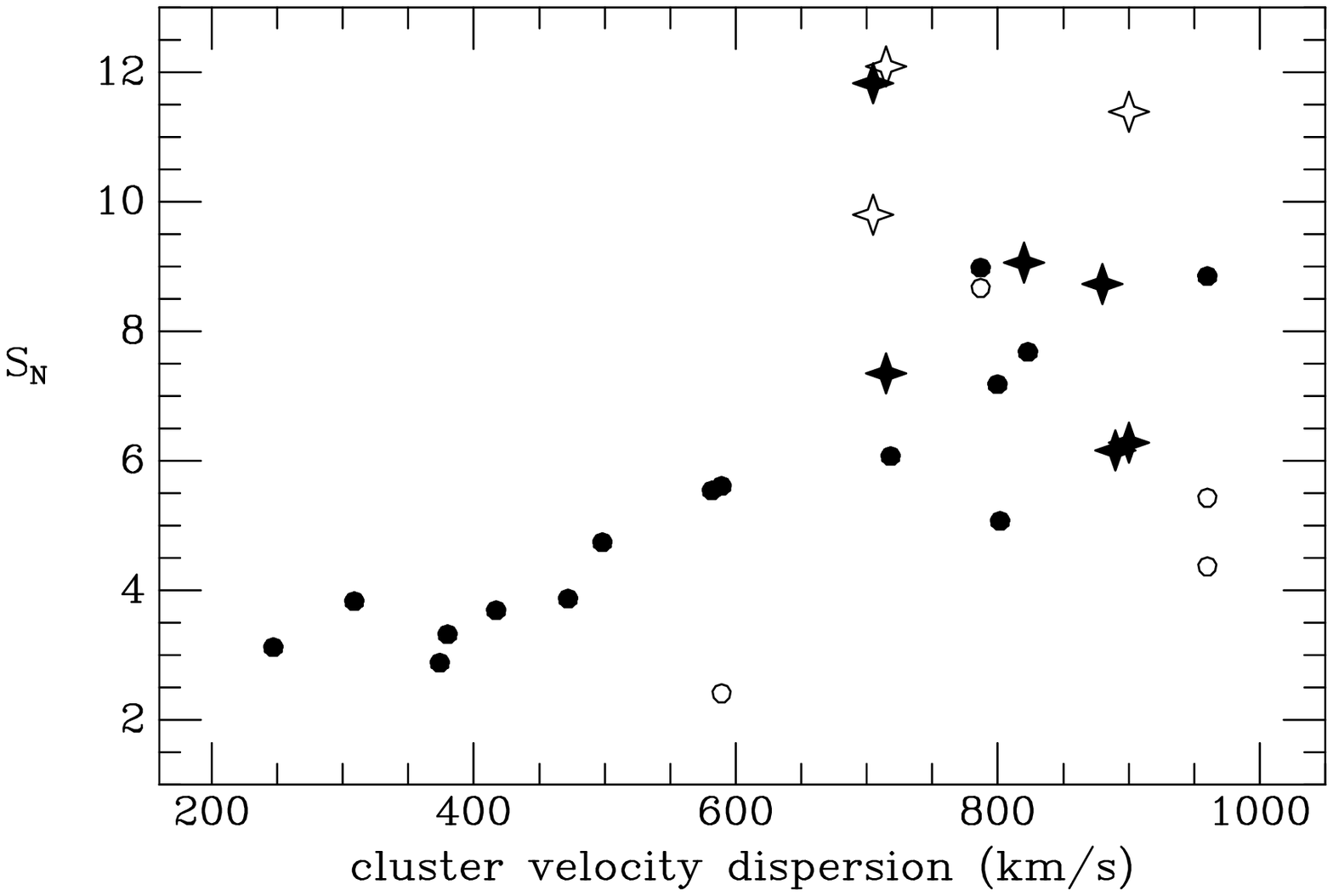, width=4.75in}}
\bigskip
\centerline{\psfig{figure=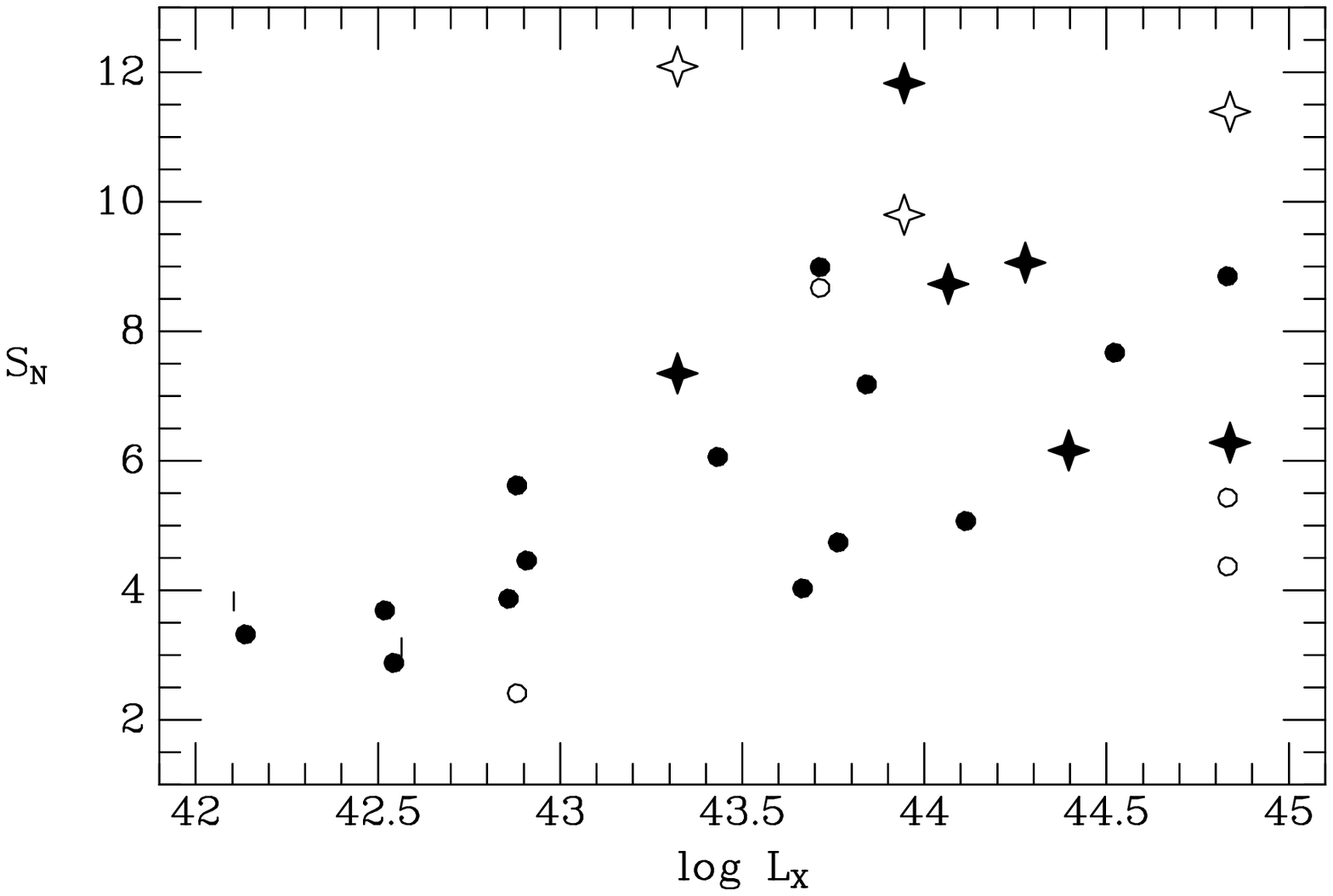, width=4.75in}}
\caption{
Correlations of the GC specific frequency $S_N$ with the velocity
dispersion (top) and the X-ray luminosity (bottom) of the host galaxy
cluster.  Filled and open circles respectively represent central and
non-central (``secondary'') bright cluster galaxies from the complete 
$c{z} < 10,000$ \kms\ sample studied by Blakeslee \etal\ (1997).  Filled
four-pointed stars show results for the cD galaxies
in the Blakeslee (1999) sample, and open stars are for several fainter
ellipticals in those same central fields.  (Short vertical
lines in the lower panel represent clusters in the former study with only
upper limits on X-ray luminosity.)
Thus, the filled symbols are all dominant central galaxies.
However, open circles are high-luminosity non-central galaxies, while
open stars are low-luminosity galaxies near the cD.  
If the number of GCs depends more on central location than on luminosity,
then the former galaxies should have lower \sn\ values than the latter
galaxies, which in turn should have \sn\ 
\hbox to 6.5in{values similar to the central
dominant galaxies, consistent with the results shown in these figures.\hfil}
\label{fig:sndens}}
\end{figure}

The data show that \sn\ for central cluster galaxies correlates well
with overall cluster properties  including velocity dispersion and
X-ray luminosity (see Figure~\ref{fig:sndens}),
as well as other measures of cluster richness.  In fact, the luminosities
of the cD galaxies show relatively little variation (e.g., Postman \& Lauer 1995),
so the correlations exhibited in  Figure~\ref{fig:sndens} are really between
the total number of GCs in the cluster center and the mass of the
cluster.  Luminous galaxies displaced from the cluster dynamical centers
do not seem to participate in these correlations (although see
Woodworth \& Harris 2000).

We have argued (Blakeslee \etal\ 1997; Blakeslee 1999) that the
observations are most easily explained if the GCs surrounding cD galaxies
formed at early times and with an approximately universal efficiency of
roughly $0.7\;$GC per $10^9\,M_{\odot}$.  However, the present-day galaxy
luminosity is fairly insensitive to this quantity, resulting in the
observed correlations of GC frequency with cluster density.  Thus, rather
than having anomalously large GC populations, the galaxies are simply
underluminous for their prominent positions in the centers of very rich
clusters.  In this ``missing light'' scenario, the gas is removed from the
central galaxy into the intracluster medium at the cluster formation
epoch, halting any subsequent star formation in the central galaxy that
would serve to lower \sn.  A similar model has been discussed by Harris
\etal\ (1998), and McLaughlin (1999) has shown that the numbers work
out in detail for the amount of gas surrounding the nearby 
galaxies M87, NGC\,1399, and NGC\,4472.

Blakeslee (1999) suggested that there may be evidence for later \sn\
evolution in these systems: the central galaxies in the more spiral-rich,
less evolved clusters had marginally higher GC specific frequencies than
those in the more centrally concentrated, spiral-poor clusters.  Again the
difference in the \sn\ values was because of the lower luminosities of the
central galaxies in the less evolved clusters, not because of larger GC
populations of in the more evolved one.  Thus, continuing star formation
and dynamical evolution in the spiral-rich clusters may allow the
\hbox{central} galaxy to grow in luminosity, without adding large numbers
of new GCs; the net result is a decrease in GC frequency.  Following
McLaughlin \etal\ (1994), who made a similar suggestion from a smaller,
much less homogeneous, dataset, we call this process ``\sn\ dilution.''
It will be interesting to see if this effect proves true; for now, the
evidence is inconclusive.~~ 

\subsection{Recent HST Data}

Of course, with \hst, it is possible to reach the GCLF turnover at the
distance of the Coma cluster and obtain a more direct estimate of \sn.  No
one has yet done a systematic study of GC populations in a large sample of
cD galaxies in rich clusters using \hst, as the required integration times
are substantial.  However, Harris \etal\ (2000) have recently used \hst\
to measure $\sn\approx4$ for NGC\,4874 in Coma.  This is significantly
lower than our value of $\sn = 9\pm2$ (Blakeslee \etal\ 1997).  There are
two basic reasons for this.  First, they assumed a distance modulus
0.35\,mag greater than ours, and this translates directly into a
$\sim35$\% difference in \sn.  Harris \etal\ point out that there is very
close agreement between the observed GC counts when compared over the same
radial and magnitude ranges.  For instance, if the value of \sn\ at
$r\sim1^\prime$ shown in Figure~\ref{fig:method} above is transformed to
their distance scale, we obtain very close agreement with $\sn\approx4$.

The more important difference stems from the radial profiles measured
in the two studies.  We found that the GC distribution was more extended
than the halo light, so that \sn\ increased with radius,
yielding a significantly higher \sn. Harris \etal\ also found a very
extended distribution in the inner parts, but then a sharp
truncation of the GC system at $r{\,\approx\,}2\,$arcmin.  It is most likely
that problems with background source contamination or the $I$-band
surface photometry at large radii (the rest of our data were all taken in
the $R$-band, where the sky is much darker and CCD's flatten better,
to avoid this problem) have affected the ground-based \sn\ measurement.
Still, the sharp truncation found near the limit of
the \hst\ data is difficult to understand.  It would be worth revisiting
this system with the next generation of \hst\ imaging cameras (ACS, WFC3)
and their much wider fields of view.

\section{Summary}

The rich globular cluster populations at the centers of galaxy
clusters constitute important probes of the central cluster potential.
In nearby clusters, they can be used to directly trace the transition
from the potential of the central galaxy to that of surrounding cluster.
In the case of NGC\,4472, the brightest Virgo elliptical, this transition
is well traced by the more spatially-extended blue GC subpopulation,
whereas the red GCs appear to share the dynamics of the stellar halo.
By obtaining large samples of GC velocities around both central and
non-central giant ellipticals, it will be possible to determine 
the mass profiles of the galaxies and constrain the slope of the dark matter 
profile near the centers of galaxy clusters.  This will provide
a useful comparison for high-resolution $N$-body simulations
(e.g., Moore \etal\ 1998).

At larger distances where GCs become too faint for spectroscopic
measurements, observations of GC populations around cD galaxies can
give a reasonable indication of the cluster core mass.  The reason
for the elevated specific frequencies of the central galaxies is
most likely that their gas was removed and their star formation halted
sometime after the formation of the GCs.  In this case, it should be
possible to observe \sn\ evolution in clusters of different
evolutionary states, but the available data are inconclusive.
As always, future observations
of GC systems from the ground and with \hst\ will be helpful in
resolving the many remaining questions.

\section*{Acknowledgments}
I thank Mike Beasley for supplying Figure\,1 and helpful comments.
This work was supported at Caltech by a Fairchild Fellowship
and at the University of Durham by a PPARC Rolling Grant in
Extragalactic Astronomy and Cosmology.

\end{document}